\documentclass[showpacs,preprintnumbers,aps,prc,floatfix]{revtex4}
\usepackage{graphicx}
\usepackage{amssymb}
\usepackage{longtable}
\begin{document}
\title
{High-spin structure and Band Termination in $^{103}$Cd} 
\author{A. Chakraborty$^{1}$, Krishichayan, S. Mukhopadhyay, S. Ray, S.N. Chintalapudi, S.S. Ghugre, 
N.S. Pattabiraman$^{2}$, and A.K. Sinha}
\affiliation
{UGC - DAE Consortium for Scientific Research, Kolkata Center,
Sector III/LB-8, Bidhan Nagar, Kolkata 700098, India}
\author{S. Sarkar$^{3}$}
\affiliation
{Department of Physics, Bengal Engineering and Science University, Shibpur, Howrah 711103, India}
\author{U. Garg and S. Zhu$^{4}$}
\affiliation
{Department of Physics, University of Notre Dame, Notre Dame, IN 46556, USA}
\author{M. Saha Sarkar}
\affiliation
{Saha Institute of Nuclear Physics, Sector-I/AF, Bidhan Nagar, Kolkata 700064, India}

\altaffiliation[$^{1}$Present address: ]
                           {{\it Department of Physics, Krishnath College, Berhampore 742101, India}} \ \\  
\altaffiliation[$^{2}$Present address: ]
                           {{\it Department of Physics, University of York, York YO10 5DD, UK}} \ \\   
\altaffiliation[$^{3}$On lien from ]
                           {{\it Department of Physics, The University of Burdwan, Burdwan 713104, India}} \ \\   
\altaffiliation[$^{4}$Present address: ]
                           {{\it Physics Division, Argonne National Laboratory, Argonne, Illinois 60439}}

\begin{abstract}
Excited states of the neutron deficient $^{103}$Cd nucleus have been investigated 
via the $^{72}$Ge($^{35}$Cl, p3n)  
reaction at beam energy of 135 MeV by use of in-beam spectroscopic methods. 
Gamma rays depopulating the excited states  
were detected using the Gammasphere spectrometer with high-fold $\gamma$-ray coincidences.
A quadrupole $\gamma$-ray coincidence analysis ($\gamma^{4}$) has been used to extend the 
known level scheme. The positive parity levels have been established up to $J = 35/2\hbar$
and $E_{x} = 7.071$ MeV.  
In addition to the observation of highly-fragmented level scheme belonging 
to the positive-parity sequences at E$_{x}\sim$ 5 MeV, the termination
of a negative-parity sequence connected by $E2$ transitions has been established at $J = 47/2 \hbar$
and $E_{x} = 11.877$ MeV. 
The experimental results corresponding to both the positive- and negative-parity sequences have been
theoretically interpreted in the framework of the core particle coupling model.
Evidence is presented for a shape change from collective prolate to non-collective oblate 
above the $J^{\pi} = 39/2^{-}$ (8011 keV) level and for a smooth termination
of the negative-parity band. 
\end{abstract}

\pacs{27.60.+j, 23.20.Lv, 21.60.Ev}

\maketitle

\section{Introduction}

	The level structures of cadmium isotopes exhibit a variety of appealing
features with the variation in the neutron number. The chain of isotopes 
with A $\leq$ 102 indicate single particle behavior resulting 
from the excitations of proton holes and the neutron particles across the 
doubly magic $Z=N=50$ core. The level structures of 
$^{98,99}$Cd, lying in the vicinity of $^{100}$Sn, have been the focus of
several experimental investigations \cite{gorska, lipo1, lipo2, blaz}. 
These studies are expected to furnish information on the single-particle energies and 
two-body matrix elements which would help develop better 
empirical interactions 
for the model valence spaces around the $N=Z=50$ magic shell closure.
While the ``fingerprint'' of core excitations or excitations involving 
1$h_{11/2}$ neutrons have been demonstrated in $^{100}$Cd from the observation 
of high-energy transitions (E$_{\gamma}$$\sim$ 2.0 -  2.6 MeV) around 
the $I = 14 \hbar$ \cite{clark1}; a theoretical prediction for the ground state of 
the nucleus $^{104}$Cd to be a superposition of prolate ($\beta_{2} \sim 0.15$) 
and oblate ($\beta_{2} \sim -0.12$) configurations 
has also been mentioned in Ref. \cite{gam}.  
With the increase in neutron number ($A = 106 - 110$), shears mechanism appears 
to develop in Cd-isotopes due to the availability 
of proton holes in the high-$\Omega$
1$g_{9/2}$ orbitals and neutron particles in the low-$\Omega$ 1$g_{7/2} / 2d_{5/2}$
and 1$h_{11/2}$ orbitals, leading to    
magnetic and anti-magnetic rotations 
 \cite{simons, chiara, kelsall, datta, clark2}.
Further, for a long time, cadmium isotopes lying close to stability line 
(111$\leq$ A $\leq$114) have been considered as examples of quadrupole 
vibrator with spherical shape \cite{comfort}.   
Also during the past decade, Cd isotopes have emerged as the
laboratory for the study of multi-phonon excitations, such as the
three-quadrupole-phonon \cite{apra, casten, lehmann} and mixed
quadrupole-octupole excitations of the type (2$^{+}$$\otimes$3$^{-}$)
\cite{gade, deepa1, garrett}. 
The observation of one-phonon mixed-symmetry state \cite{deepa2} 
also adds to the diversity of excitation modes in Cd isotopes. 

	The $^{103}$Cd nucleus, with two proton holes and five valence 
neutrons, with respect to the $^{100}$Sn core, lies in the upper part of the 
1$g_{9/2}$ proton sub-shell and in the lower part of the $gds$ 
neutron sub-shell. 
This is a transitional region between the spherical nuclei close 
to the doubly magic nucleus $^{100}$Sn and the deformed nuclei with 
more valence particles. The prolate deformation-driving character of 
the low-$\Omega$ $h_{11/2}$ orbitals, however, allows the formation of 
quasi-rotational bands in such nuclei. The relatively small number 
of available valence particles implies that these quasi-rotational bands would 
terminate at rather low spins that can be accessed experimentally. 
	
	Excited states in $^{103}$Cd have been previously investigated  
by several groups \cite{her,meyer,palacz}. The first excited 
state in $^{103}$Cd with $J^{\pi}$ = 7/2$^{+}$ at $E_{x}$ = 188 keV was 
established by Lhersonneau {\it et al.} \cite{her} following 
the $\beta$-decay of $^{103}$In. Two high-spin studies have 
also been reported in the subsequent years \cite{meyer,palacz}. 
The present work is aimed at extending the available information on the 
excited states of $^{103}$Cd up to the energy and spin regime where 
one would expect termination of bands, and also to explore
the interplay between the single particle and deformed structures. 
Apart from a significant extension of the positive-parity
level sequences up to $J^{\pi} = (35/2^{+})$ and $E_{x} \sim$ 7 MeV,
the present investigation has furnished evidence for
a smooth termination of the negative-parity decoupled band. The 
phenomenological core particle coupling model (CPCM) has been employed to
understand the microscopic properties of the 
level structures and other spectroscopic observables.

\section{Experimental Procedure and Data Analysis}

	The experiment was performed with  
the $^{72}$Ge($^{35}$Cl, p3n)$^{103}$Cd reaction, using a 
135-MeV $^{35}$Cl beam provided by the ATLAS facility
at the Argonne National Laboratory.  
The $^{72}$Ge target (of 1 mg/cm$^{2}$ thickness) was evaporated 
onto a 15 mg/cm$^{2}$-thick gold foil; 
the choice of backing provided sufficient stopping power to slow 
down and stop the recoiling nuclei produced in this reaction. 
A thin (40 $\mu$g/cm$^{2}$) 
Al layer was evaporated between the target and the backing
to avoid migration of target material into the backing.  

	The de-exciting $\gamma$-rays were detected using the 
Gammasphere spectrometer \cite{nolan,janssens,lee} 
in the stand alone mode. 
Events were written on magnetic tape when at least three suppressed
Ge detectors in the array detected $\gamma$ rays within the prompt 
coincidence time window. A total of approximately 2$\times$10$^{9}$ 
``triple coincidence'' events were recorded during the experiment.
The coincidence events were sorted into a $\gamma$$^{4}$ 
histogram and asymmetric angle - selected matrices (see below) 
for off-line analysis. The data analysis was performed using the
Linux based software packages IUCSORT \cite{nsp1,nsp2} and RADWARE \cite{rad}.  

	The other strongly populated nuclei with A$\sim$100 
in the present experiment are $^{100,101}$Pd, $^{103,104}$Ag
and $^{104}$Cd. The high-fold coincidence data 
helped us to unambiguously place in the level 
scheme of $^{103}$Cd some transitions that are identical  
in energy to those from the aforementioned neighboring evaporation residues. 

	Multipolarities of emitting $\gamma$-rays from $^{103}$Cd
were deduced from the analysis of Directional Correlations of
$\gamma$-rays de-exciting Oriented states (DCO ratio method) 
\cite{kramer}. 
The method is based on the observed coincidence intensity
anisotropy, obtained from the angle dependent $\gamma - \gamma$
coincidence. This anisotropy ratio is denoted by $R_{DCO}$.
In order to
extract DCO-ratios, the coincidence data were sorted into
an asymmetric matrix whose one axis corresponded to the
$\gamma$-ray energy deposited in the detectors at 35$^{\circ}$
and 145$^{\circ}$ and the other axis corresponded to the
$\gamma$-ray energy deposited in the detectors at 90$^{\circ}$.
The DCO-ratios are, then, defined as:

$$
R_{DCO} = \frac{ I_ {{\gamma}_1}~ at ~  35^{\circ} and ~  145^{\circ};
~ gated~  with~ \gamma_2~ at~  90^{\circ} }
{I_ {{\gamma}_1}~ at ~ 90^{\circ}; ~ gated~ with ~\gamma_2
~ at ~  35^{\circ} and ~  145^{\circ}}
$$                                                                                            

	The DCO ratios were at first calibrated by measuring 
several known $E2$ and dipole transitions in $^{100}$Pd \cite{zhu} 
and $^{101}$Pd \cite{popli}, which were also 
populated in the present experiment. 
Gating on a stretched quadrupole transition $\gamma_{2}$,
one obtains value of $R_{DCO}$ $\sim$1.0 
for a stretched quadrupole transition,
and $\sim$0.5 for a 
stretched pure dipole. By gating on a stretched pure dipole
transition, one would get $\sim$2.0 for a quadrupole and $\sim$1.0
for a pure dipole transition. For mixed dipole/quadrupole $\gamma_{2}$
transitions, $R_{DCO}$ depends on the value of the mixing ratio.

	In assigning spins and parities to the observed levels, we have assumed
that the stretched quadrupole transitions are of $E2$ multipolarity (in general),
the pure dipoles can have either E1 or M1 character, and 
the mixed transitions are of M1/E2 type. Although the DCO-ratios
can not distinguish between stretched E2's and $\Delta$I = 0 dipoles
or certain $\Delta$I = 1 admixtures, the spin
assignments are relatively unambiguous since the $^{103}$Cd level scheme is
quite complex with many parallel decay pathways and many cross-over
connections.
Here, we have used the general yrast
argument that levels populated in heavy-ion reactions usually
have spins increasing with increasing excitation energy. 

        Although for most of the transitions, the DCO-ratios have been
obtained using the gating transition with known multipolarity, 
the same procedure could not be utilized to determine the DCO-ratios
for the weak transitions lying above $\sim$ 5 MeV excitation.
The multipolarities of such transitions have been
inferred from the measured DCO-ratios of the known intense
transitions, lying below $\sim$ 5 MeV excitation, with the corresponding
weak transition used as the gating transition in the DCO analysis.

\section{Experimental Results}

\begin{figure*}
\includegraphics{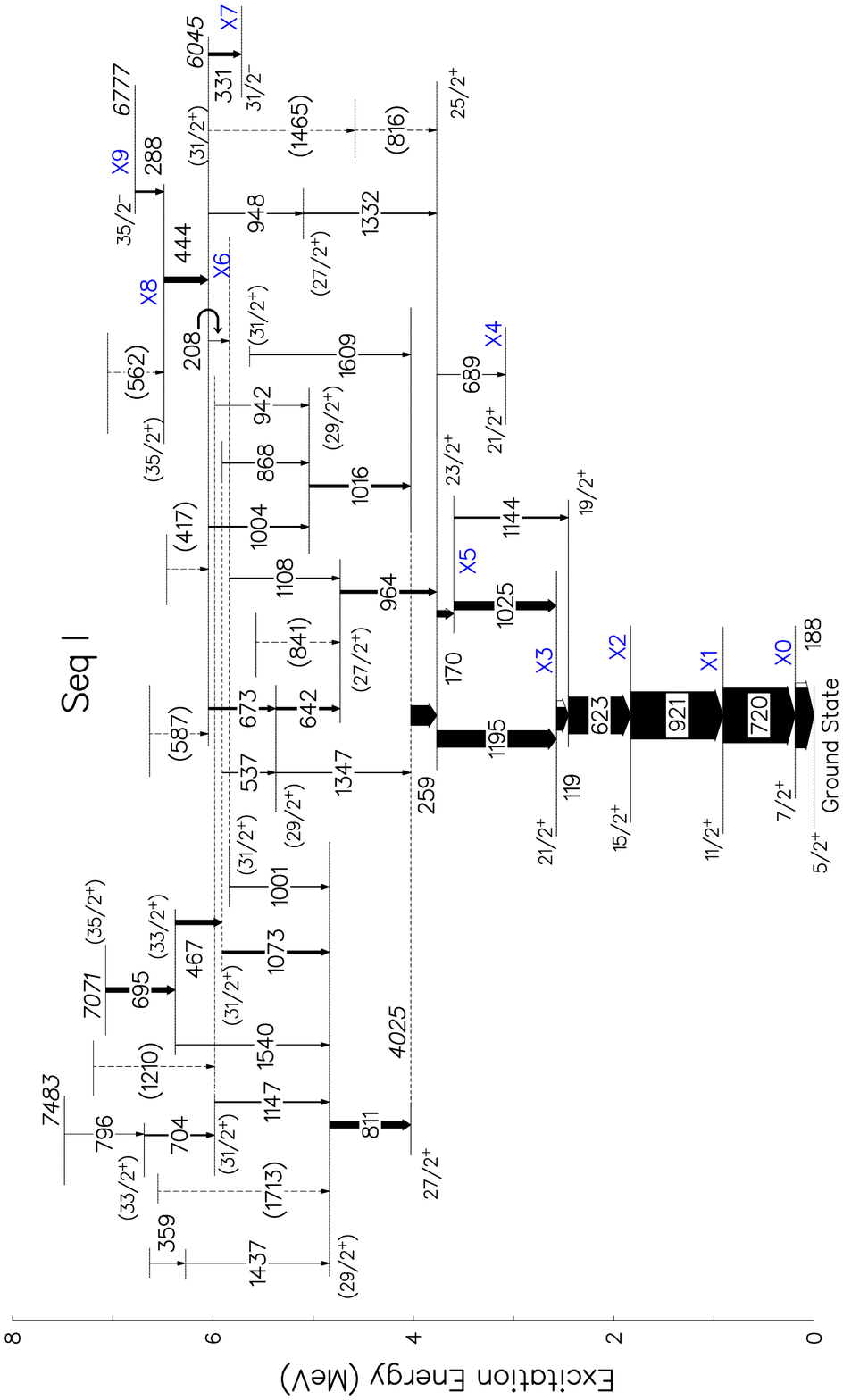}
\caption{
\label{fig:levelscheme1}
``(Color online)'' Part of the proposed level scheme, including $Seq$ I, 
for $^{103}$Cd as obtained from the 
present work. Widths of arrows connecting the levels are approximately 
proportional to the relative $\gamma$-ray intensities. 
Transition energies and the energies of levels (where shown)
are in keV. The levels common to different sequences
have been denoted as X$_{i}$'s ($i$ = 0,1,2,...,9) to guide the eye.
}
\end{figure*}

\begin{figure*}
\includegraphics{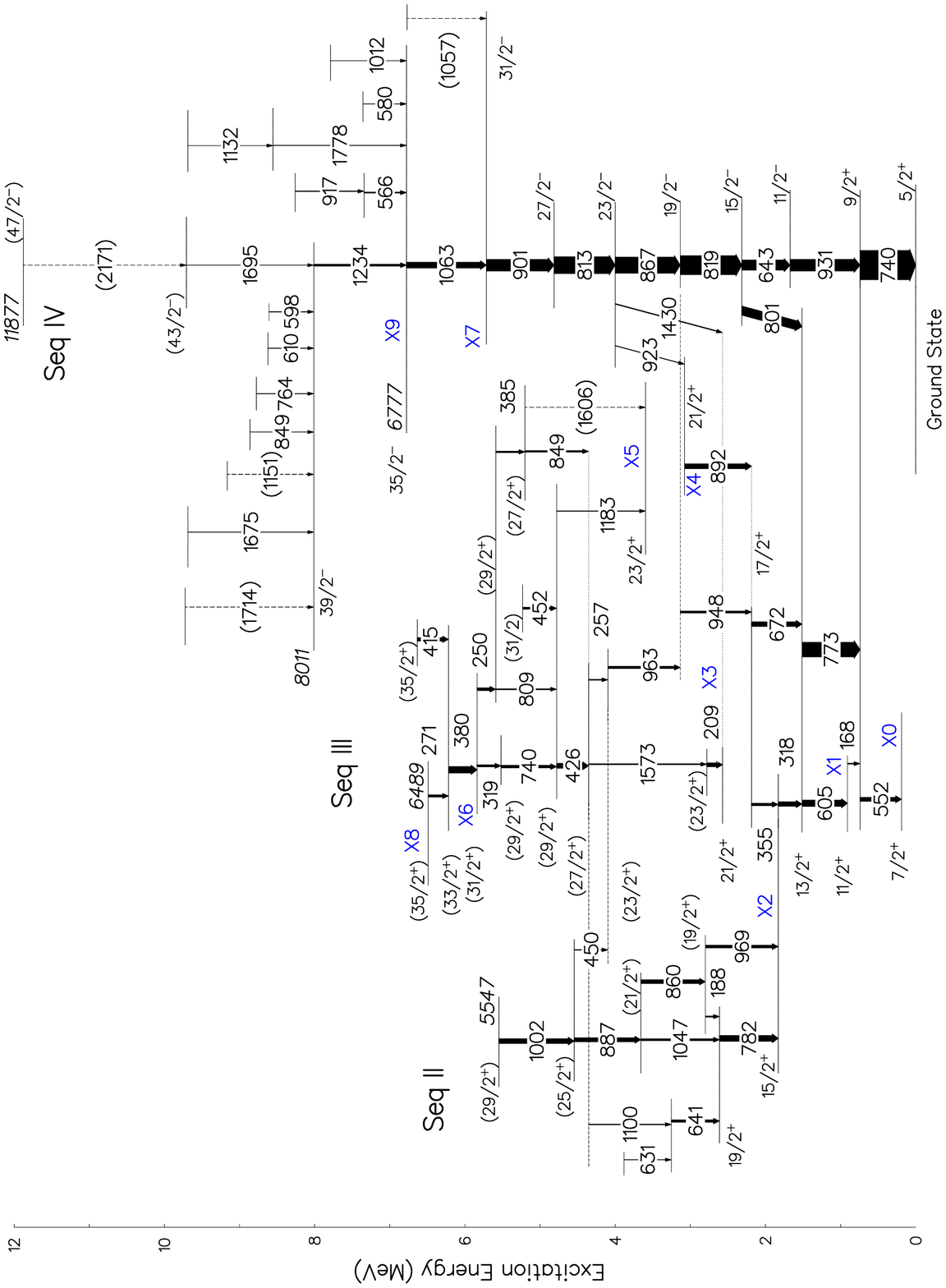}
\caption{
\label{fig:levelscheme2}
``(Color online)''Part of the proposed level scheme, including $Seq$ II, III,
and IV, for $^{103}$Cd as obtained from the
present work. Widths of arrows connecting the levels are approximately
proportional to the relative $\gamma$-ray intensities.
Transition energies and the energies of levels (where shown)
are in keV. The levels common to different sequences
have been denoted as X$_{i}$'s ($i$ = 0,1,2,...,9) to guide the eye. 
}
\end{figure*}

	The $^{103}$Cd level scheme resulting from the analysis of the 
coincidence $\gamma^{4}$ histogram
is shown in Figs.~\ref{fig:levelscheme1} and ~\ref{fig:levelscheme2}. 
In the figures, the energies of the $\gamma$
transitions are labeled in keV and the widths of the arrows are approximately proportional
to the transition intensities. 
The main positive-parity sequence has been extended up to an excitation energy of $\sim$ 7 MeV 
(with $J^{\pi}$ = (35/2$^{+}$))
and the negative-parity sequence is extended up to $\sim$ 12 MeV
(with $J^{\pi}$ = (47/2$^{-})$). About 50 new transitions have been placed in the level scheme.
Several level sequences have been labeled (I,II,III,IV) 
to facilitate further discussion.

\begin{figure*}
\includegraphics{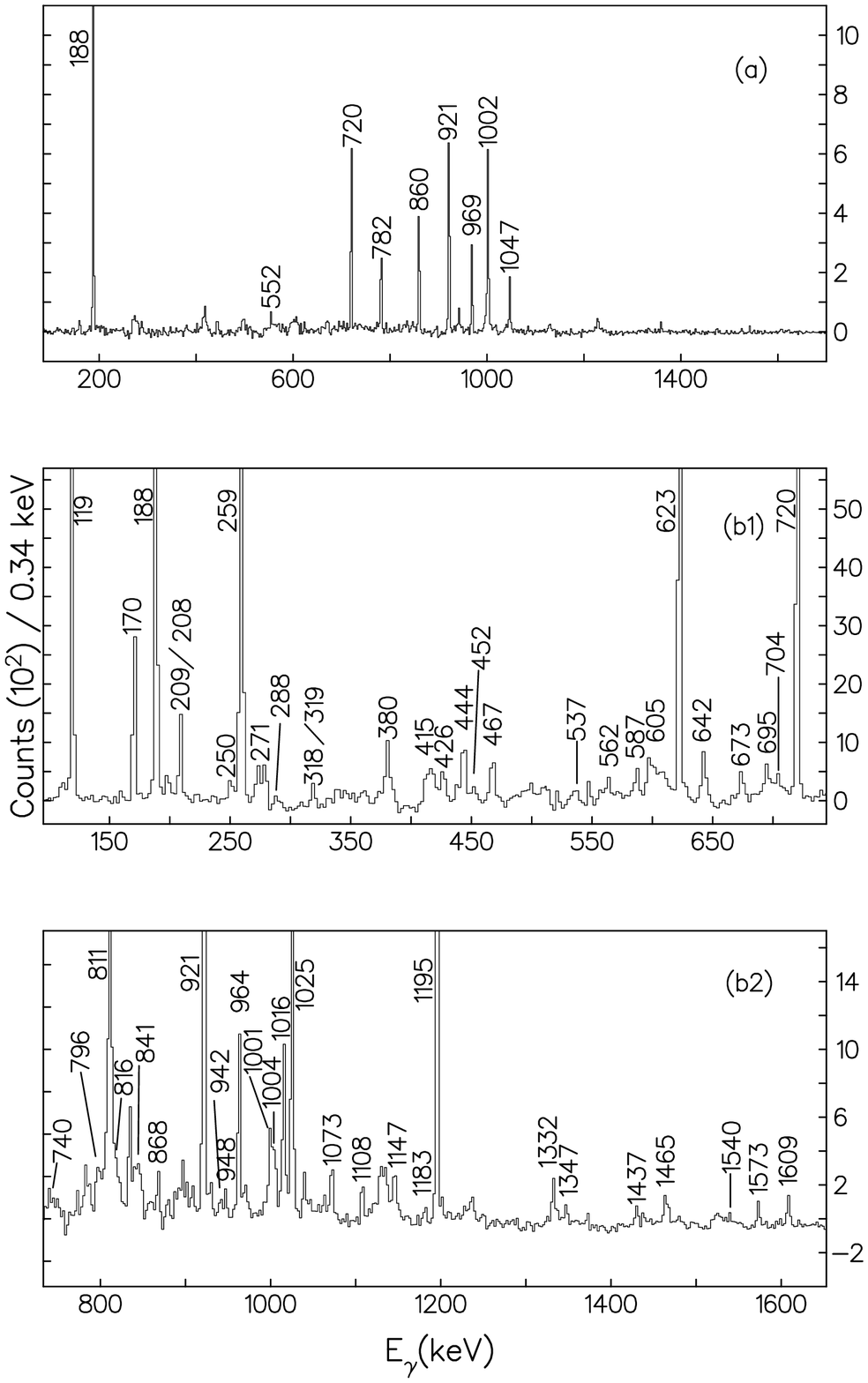}
\caption{
\label{fig:spectra_p}
(a) Representative $\gamma\gamma\gamma\gamma$ coincidence spectra of the 887-keV
transition with other members of $Seq$ II. 
(b1-b2) Coincidence spectra for the transitions belonging to $Seq$ I. 
Peaks labeled with their energy values (in keV) have been assigned to $^{103}$Cd. 
}
\end{figure*}

\begin{figure*}
\includegraphics{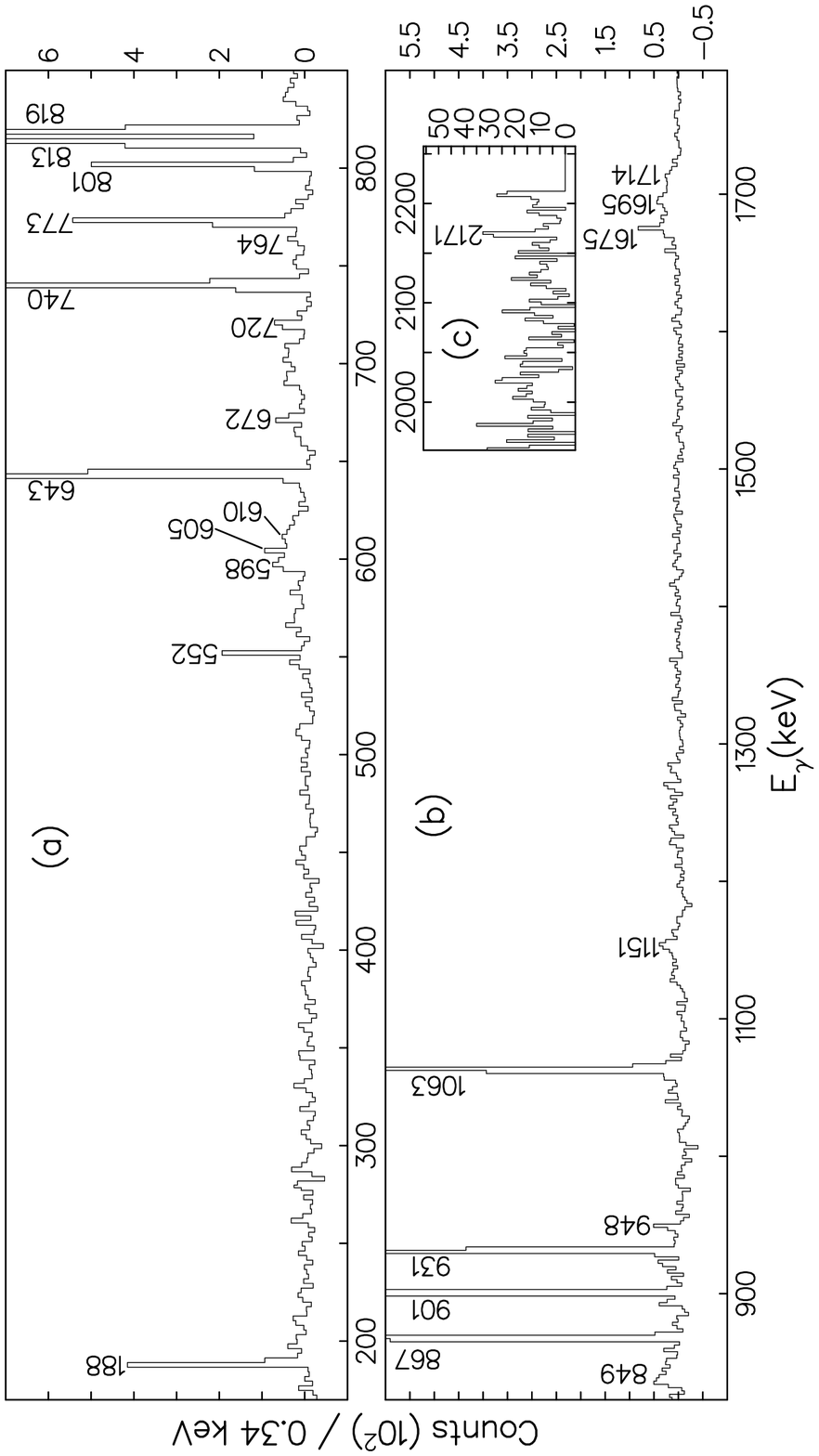}
\caption{
\label{fig:spectra_n}
(a-b) Representative $\gamma\gamma\gamma\gamma$ coincidence spectra of the 1234-keV
transition with other members of $Seq$ IV. 
Peaks labeled with their energy values (in keV) are assigned to $^{103}$Cd. The spectrum in the high energy
inset (c) was obtained from the sum of single-gated $\gamma$-ray spectra for 
the 901-,1063-, and 1234-keV transitions.
}
\end{figure*}

	Typical prompt $\gamma\gamma\gamma\gamma$ coincidence spectra for positive-parity
levels in $^{103}$Cd are shown in Fig.~\ref{fig:spectra_p}.
Fig.~\ref{fig:spectra_n}(a-b)
depicts the representative $\gamma$$^{4}$ coincidence spectrum for
the transitions belonging to negative parity sequence ($Seq$ IV). 
The newly observed weak transitions feeding
the 39/2$^{-}$ level at 8011 keV are clearly seen in the figure. 
The presence of the top most 2171-keV transition of $Seq$ IV could be
identified from a sum of single-gated $\gamma$-ray coincidence spectra 
with the gates set on 901-,1063-, and 1234-keV transitions 
(see Fig.~\ref{fig:spectra_n}(c)).
A similar procedure has been adopted 
in Refs. \cite{lafosse,schnare} to assign weak transitions
belonging to the corresponding level schemes.
However, Fig.~\ref{fig:spectra_n}(b)
does not indicate the presence of any transition with energy more than 1.6 MeV;
the 2171-keV transition has, therefore, been placed in the level scheme only as
tentative.

	The main positive-parity sequence ($Seq$ I) is built on the 5/2$^{+}$ ground state.
Both positive-parity sequences ($Seq$ I and III) exhibit an 
irregular and complex structure above 4 MeV, a feature typical of 
particle-hole excitations in nearly-spherical nuclei.  
The negative-parity sequence ($Seq$ IV) built on the 11/2$^{-}$ bandhead at $E_{x}$ = 
1.671 MeV exhibits a regularity up to $J^{\pi}$ = 39/2$^{-}$ and $E_{x}$ = 8.011 MeV.
Above this state, a certain regularity persists, though weakly, up to $J^{\pi}$ $\sim$ $(47/2^{-})$ at
$E_{x}$ = 11.877 MeV. However, the level structure appears to gradually become irregular around
$E_{x}$ $\sim$ 6.8 MeV, likely indicating a change in structure. 

	Numerous cross-over transitions, established from the observed
coincidence relations, provide checks for the placement and ordering of transitions 
in many cases in the complex scheme. 
The placement of all the known transitions is in agreement with the previous work \cite{palacz}. 
In addition to the extension of the existing level scheme, the present investigation also leads to
the unambiguous placement of some of the transitions, which were tentatively placed in the work of Palacz 
{\it et al.} \cite{palacz}. As there are a number of doublets present in the level scheme, different gates 
have been used to determine their relative intensities and DCO-ratios.  

	The 1128-keV transition, which appears to belong to $Seq$ I could not be convincingly 
placed in the level scheme. Also the 262- and 373-keV transitions, which appear to be in coincidence with
the 1573-keV and other transitions in the cascade, could not be placed in the level scheme due to the poor 
statistics. Furthermore, the 1040-keV transition could not be placed 
in the level scheme although it appears to be in strong coincidence with 259-keV and the other low-lying
transitions of $Seq$ I. There seems to be a linking between the 1234-keV ($Seq$ IV)
and 1609-keV ($Seq$ I) transitions; however, the coincidence relationship between the two
could not be established from the present data. 

	The $\gamma$-ray energies, intensities, DCO-values, and multipolarities for the transitions 
assigned to $^{103}$Cd have been summarized in Table I. The uncertainties in the $\gamma$-ray energies
correspond to the error due to peak fitting.
The uncertainties in the intensities encompass errors
due to background subtraction, peak fitting, statistical fluctuations, and efficiency correction; 
the error bars may have been somewhat underestimated for the weakest transitions.
The uncertainties quoted in the DCO-ratios include the errors due to background subtraction,
statistical fluctuations, and peak fitting.

	Although the intensity balance between the transitions feeding to and decay out from
a particular level is equal within error for most of the cases, there is a
marked deviation observed for the level with $J^{\pi}$=19/2$^{+}$. 
This level was found to be isomeric in the work of Palacz {\it et al.}
 \cite{palacz}. 
A large intensity imbalance between feeding and decay has similarly been observed 
for the 27/2$^{+}$ level at 4025 keV.   

\section{Theoretical Calculations and Discussion}

	The experimental level 
structure of $^{103}$Cd (see Figs.~\ref{fig:levelscheme1} 
and ~\ref{fig:levelscheme2}) consists of 
three positive-parity sequences ($Seq$ I - III) and a negative-parity 
band ($Seq$ IV) built on $J^{\pi}$ =  11/2$^{-}$ state. 

Meyer {\it et al.} \cite{meyer} suggested that the low-lying yrast 
positive-parity levels  
are the members of a decoupled
band, built on the first excited 7/2$^{+}$ level, 
with the odd $\nu(1g_{7/2}$) quasi-particle coupled
to the core. This was based on the similarity of the
excitation energies of the 11/2$^{+}$, 15/2$^{+}$, 19/2$^{+}$, ......
levels in $^{103}$Cd to the 2$^{+}$, 4$^{+}$, 6$^{+}$, .....
levels of the respective ground state bands in $^{102,104}$Cd.
For a complete decoupled band \cite{stephens} built on 7/2$^{+}$ band
head, the excitation energies of the levels belonging to
the sequence (with signature, $\alpha$ = -1/2) 
11/2$^{+}$, 15/2$^{+}$, 19/2$^{+}$,
....... should be lower than their 
respective unfavored signature ($\alpha$ = +1/2) 
partners 9/2$^{+}$, 13/2$^{+}$,
17/2$^{+}$, ....... . The levels corresponding to 
the unfavored signature partners are populated weakly in
heavy-ion reactions and in many cases can not be observed 
experimentally.
As Meyer {\it et al.} \cite{meyer} had not observed the 
unfavored signature partners, it supported their conclusion 
for considering the positive parity band  
with 7/2$^{+}$ band head as the decoupled band.

In the subsequent work of Palacz
{\it et al.} \cite{palacz}, as well as in the present investigation,
the unfavored signature partners have been observed.
It is found that although the signature splitting
is strong, the unfavored signature partners are still energetically
lower than the corresponding next favored ones
and hence the band is not completely decoupled. 

The experimental level energies of the negative-parity band
are also found to be in close agreement with those of the corresponding 
levels of the ground state bands of $^{102,104}$Cd.
This suggests that the excited negative-parity levels in $^{103}$Cd
originate essentially due to the coupling of the 
1$h_{11/2}$ neutron quasiparticle with the collective
excitations of the $^{102}$Cd or $^{104}$Cd core.
As such, the observed negative-parity band is 
identified as a decoupled band arising from $\nu h_{11/2}$
Nilsson orbits with the Fermi level lying near
a low-$\Omega$ orbital ($\Omega$ = 1/2 in the present case).

From the shell model point of view, $^{103}$Cd nucleus
can be thought of as having two proton-holes $\pi(1g_{9/2})^{-2}$
and five neutrons in the $\nu(2d_{5/2},1g_{7/2},3s_{1/2},2d_{3/2},
1h_{11/2})$ valence space with respect to doubly magic
$^{100}_{50}Sn_{50}$ core. The 5/2$^{+}$ ground state and the
low-lying positive parity excited states up to 21/2$^{+}$
can be built from the [$\pi(g_{9/2})^{-2} \otimes \nu(d_{5/2})^{-1}$]
multiplet. Beyond 21/2$^{+}$, neutron excitations to 1$g_{7/2}$
orbital are essential. This can provide a maximum $J^{\pi}$ of 
39/2$^{+}$. The maximum $J^{\pi}$ that can be built in the 
[$\pi(g_{9/2}), \nu(gdsh)$] valence space is 55/2$^{+}$ which
comes from the
[$\pi(g_{9/2})^{-2} \otimes \nu(g_{7/2}h^{4}_{11/2})$]
configuration. Observed maximum spin is then, 35/2 
for the positive parity. There are a large number of
configurations belonging to the partitions 
[$\pi(g_{9/2})^{-2} \nu(gdsh)^{5}$]. Most of the states
can be built in a variety of ways. Therefore, the underlying
structure of most of the levels is expected to consist
of admixture of configurations of many multi-particle-hole
excitations. This gives rise to a complex and irregular 
structure beyond the first 21/2$^{+}$ state belonging to 
the energetically lowest configuration 
[$\pi(g_{9/2})^{-2} \otimes \nu(d_{5/2})^{-1}$].
For the positive parity levels with 
$J^{\pi}$ $>$ 21/2$^{+}$, 
$\nu(d_{5/2})^{-1} \rightarrow \nu(gdsh)$ excitations
are needed and they appear at $\geq$ 4 MeV.
This can be estimated from the consideration of the 
single particle energies \cite{palacz} of the valence orbitals
and the expected configuration mixing consisting of energetically
higher lying configurations. This qualitative picture for
the positive parity levels has actually been observed in the
shell model calculations by Palacz {\it et al.} \cite{palacz}.   
They showed that the positive-parity levels
were highly configuration mixed. Further, $\beta_{2}$ values  
of even Cd-isotopes with $A \geq 102$ are known
to lie in the range 0.17 - 0.19  \cite{raman}. 
Thus, one might expect a small deformation in $^{103}$Cd and, consequently,
the onset of mild collectivity. 
The regularity observed in the negative-parity band could not 
be reproduced in shell model calculations even with large valence 
basis space considered by Palacz {\it et al.} \cite{palacz}. 

The possibility of a mild deformation in $^{103}$Cd makes 
the Core Particle Coupling Model (CPCM), proposed by
Muller and Mosel \cite{muller,maitrayee}, an ideal
choice to interpret the observed excitation modes. 
The CPCM employed here has the added advantage
that the experimental level energies of the core can
be directly used as inputs. As the core simulation is not needed, 
the approach is especially suitable for very weakly deformed
systems where variable moment of inertia (VMI) or constant
moment of inertia (CMI) approaches within the standard particle
rotor model fail due to the non-rotational nature of the core.  

	The CPCM calculations were performed for the positive 
as well as negative-parity energy levels using both $^{102}$Cd 
and $^{104}$Cd cores. 
Similar energy eigenvalues were obtained for most of the
levels in $^{103}$Cd with both the cores. This was possible
with a slight
adjustment of the attenuation factor of the Coriolis 
matrix element, keeping all other parameters
the same.
But the level energies as well as the transition 
probabilities of the higher-spin levels appear to be better
reproduced with the $^{102}$Cd core and, hence, only the results of the 
calculations performed with the $^{102}$Cd core are being presented.

	All the parameters used for the CPCM calculation for the
positive parity levels are listed in Table II. 
The Nilsson single particle orbitals for the neutron {\it viz.} 
3$s_{1/2}$, 2$d_{3/2}$, 2d$_{5/2}$, 1$g_{7/2}$, and 1$g_{9/2}$, 
belonging to N = 4 oscillator shell were considered for the calculation 
of positive-parity levels. 
The neutron Fermi level lies near the 3/2$^{+}$[411] and 5/2$^{+}$[413] 
Nilsson orbitals, originating predominantly from 1$g_{7/2}$ and 2d$_{5/2}$, 
respectively. 
The pairing gap parameter $\Delta$ was estimated from the experimental 
odd-even mass difference and is similar to the value obtained from the 
expression $\Delta$ = 12/A$^{1/2}$. 
It is generally observed that the use of
$\Delta$ value obtained from the experimental odd-even mass difference
results in an overestimation of the Coriolis interaction, requiring 
comparatively larger Coriolis attenuation coefficients (see Table II). 
The experimental core ($^{102}$Cd) energies from
0$^{+}$ to 14$^{+}$ states \cite{lieb} were supplied as inputs.
Since the experimental energies in $^{102}$Cd are available only up 
to $J^{\pi}$$ = $14$^{+}$, the energies of the higher lying levels 
(16$^{+}$ - 30$^{+}$) of the core were obtained from a fit 
[$E(R)$ = a + b$R$ + c$R^{2}$] to the low-lying known levels 
and were 
also fed in as inputs; here, $R$ denotes the spin of a particular 
level of the core and $E(R)$ is the energy of the corresponding level.  
The large b/c ratio (=31.6), obtained from the fitting,
and the value of 
$R_{4}$(= ${E}_{{4}^{+}}$ / ${E}_{{2}^{+}}$ = 2.1) 
are indicative of a vibrational nature of the core.    

	The $M1$ transition rates have been 
calculated using $(g^{s}_{\nu})_{free}$ = -3.826, $(g^{l}_{\nu})$ = 0.0, 
and $g_{R} (= Z/A_{core})$ = 0.47. 
For the calculation of $B(E2)$ values, the 
multiplicative factor for the effective charge for the
qasineutron is 0.5 \cite{sudeb}.
The contribution of the core is included through
the intrinsic quadrupole moment $Q_{0}$ using the
relation given in Ref. \cite{raman}.

	The negative-parity levels have been calculated using the same Nilsson 
parameters as those used for the positive-parity levels with the attenuation of the 
Coriolis matrix elements by a factor of 0.97. 
Only the neutron Nilsson orbitals originating from the spherical 
single particle 1$h_{11/2}$ orbital, belonging to $N$ = 5 major shell, were 
included in our calculations. 
Calculations were also carried out for the case of
oblate deformation ($\delta$ = -0.14), keeping all other parameters
unchanged. The results of all these calculations are discussed below. 

\subsection{Positive-parity Sequences}

\begin{figure*}
\includegraphics{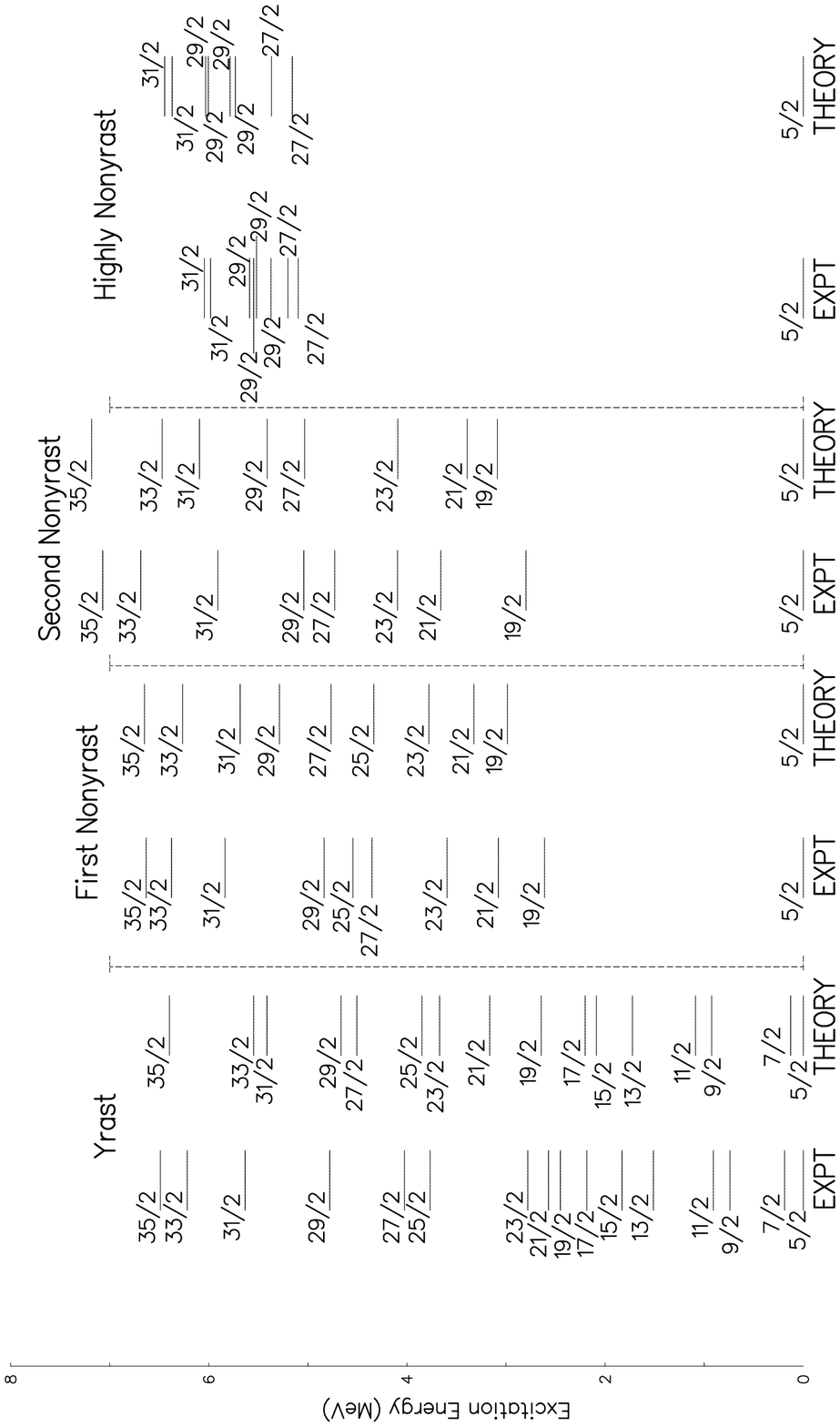}
\caption{
\label{fig:levels}
Comparison of the positive-parity levels of $^{103}$Cd with the predictions of CPCM 
calculations; see text for details. 
}
\end{figure*}

	Theoretical energy eigenvalues for the
positive-parity states obtained from CPCM calculation 
are compared with the experimental results in 
Fig.~\ref{fig:levels}. The calculations successfully reproduce 
the observed ordering, energy spacing and the signature splitting
of the levels. 

	The 5/2$^{+}$ ground state is found to be dominated
by the Nilsson basis states 3/2[422] (= 76$\%$ $\nu$ $d_{5/2}$)
and 5/2[413] (= 86$\%$ $\nu$ $d_{5/2}$) contributing 
about 18$\%$ and 67$\%$, respectively. Basis states involving $\nu$ $g_{7/2}$
also have non-negligible contributions; the contributions from 
the 3/2[411](= 81$\%$ $\nu$ $g_{7/2}$) and 1/2[420] (= 55$\%$ $\nu$ $g_{7/2}$) 
Nilsson states are about 4.4$\%$ and 4$\%$, respectively. 
The structure of the other
members of the yrast (+,+1/2) group of 
$\Delta$I=2 levels (except the 9/2$^{+}$ state) are mostly dominated by 1/2[420] and 3/2[411]
basis states and the 13/2$^{+}$ state is an almost pure 3/2[411] state. 
The yrast 9/2$^{+}$ state, the first excited state of this group, 
appears to be highly mixed with sizeable contributions 
from the close-lying 5/2[413]( 20$\%$), 3/2[422](30$\%$), 1/2[420](12$\%$)
and 3/2[411](20$\%$) $\nu$ $g_{7/2}$ orbitals.  

	The yrast 7/2$^{+}$ level originates predominantly
from the 1/2[420] (29$\%$) and 3/2[411] (46$\%$) orbitals, with admixtures from 
5/2[402](= 89$\%$ $\nu$ $g_{7/2}$)(8.41$\%$), 1/2[411](= 30$\%$ $\nu$ $g_{7/2}$)
(7.29$\%$) and 1/2[411](= 38$\%$ $\nu$ $s_{1/2}$)(7.29$\%$).
Other members of the yrast (+,-1/2)
group of levels up to 35/2$^{+}$ (except the 11/2$^{+}$ level) have a major contribution from 
3/2[411] orbital; the 11/2$^{+}$
level originates predominantly from 3/2[411](58$\%$) and 1/2[420](26$\%$).

It is interesting to understand the significance of the
agreement of the CPCM results with the experiment. The
positive parity levels which are supposed to originate
from the admixture of a large number of configurations
belonging to [$\pi(g_{9/2})^{-2}\nu(gdsh)^{5}$] partitions
can in effect be decomposed into collective excitations
(of $^{102}$Cd) based on the configuration mixing of
states belonging to [$\pi(g_{9/2})^{-2} \otimes \nu(gdsh)^{4}$]
configurations of the $^{102}$Cd core, coupled to a neutron
in $gds$ orbitals. 

The model wavefunctions were used to
calculate electromagnetic transition rates.
The experimental and theoretical branchings
for the levels have been compared 
in  Table III. It may be noted that the trend of
the observed branchings for most of the levels
could be reproduced in the present calculations.
This validates the wavefunctions obtained in the
calculations and further corroborates the
underlying structure assumption of the levels,
{\it i.e.,} a neutron quasiparticle coupled to the
collective excitation modes of the
$^{102}$Cd core. 
The discrepancy between the predicted
and experimental values of the branching ratios
for levels with $E_{x}$ $\geq$ 4 MeV might be attributable 
to the unobserved transitions
feeding the 27/2$^{+}$ level at 4025 keV;
a large intensity imbalance
between the feeding and decay pattern for this
level supports this conjecture.

\subsection{Negative-parity Sequence}
The decay of the negative parity states, belonging
to $Seq$ IV, is dominated by sequences of $E2$ transitions
between rather regularly spaced levels. The onset
of this regular pattern can be traced to the neutron excitation
to the $\nu h_{11/2}$ orbital, which is mandatory
to generate the observed negative parity levels.
The shape driving properties of the intruder
$\nu h_{11/2}$ orbital are obvious with the appearance
of regularity in the level scheme at relatively low excitation.

The negative-parity levels
up to $J^{\pi}$ = 39/2$^{-}$ are connected
by strong interband transitions. 
However, above the
$J^{\pi}$ = 35/2$^{-}$ level at $E_{x}$ = 6777 keV,
the level scheme exhibits an irregular pattern with
several parallel feeding transitions. These
transitions are extremely weak. The remarkable
dissimilarity in levels above and below the 
$J^{\pi}$ = 35/2$^{-}$ state is suggestive of
a change in the level structure at high spins.
The fragmentation of intensity into parallel cascades
above the 35/2$^{-}$ state could be attributed to the
contribution of other particle-hole excitations.
This conjecture, based on the phenomenological observations, 
is also corroborated by the CPCM and TRS calculations (discussed
below in this section).  

The negative-parity band persists, although weakly, 
up to the (47/2$^{-}$) level at 11877 keV. 
If the negative-parity band is based on the 
$\pi(g_{9/2})^{-2} \otimes \nu(g_{7/2}d_{5/2})^{4}(h_{11/2})^{1}$
configuration,
this is expected to generate a maximum
angular momentum of 47/2$^{-}$, which, indeed,
corresponds to the highest observed spin in 
this band. This may be understood as the
termination of the band at $J^{\pi}$ = (47/2$^{-}$).  

\begin{figure*}
\includegraphics[height = 20 cm]{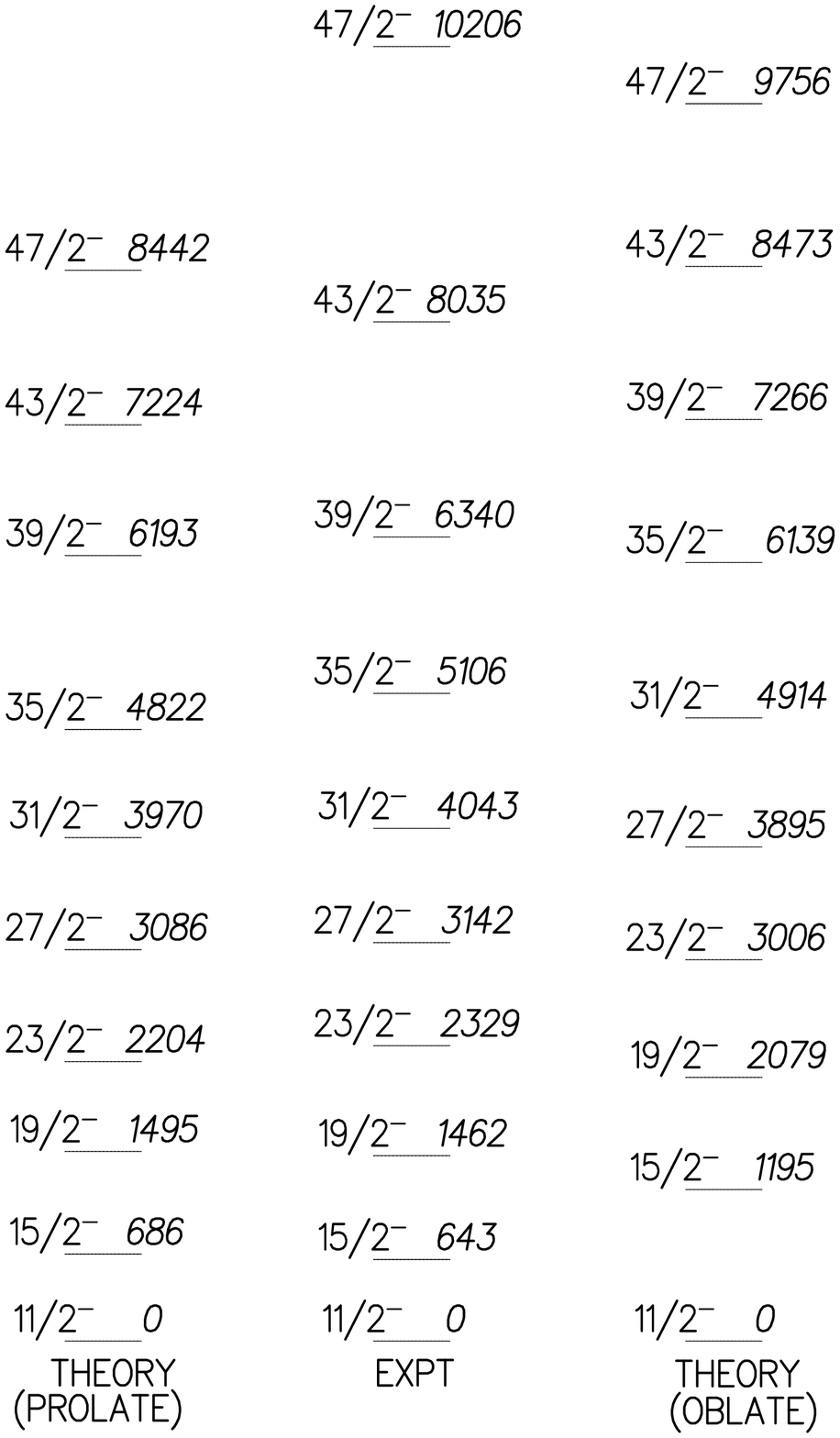}
\caption{
\label{fig:prolate+oblate}
Comparison of the experimental and theoretical excitation energies of negative-parity 
levels in $^{103}$Cd. 
Energies of the levels are labeled in keV. The calculations have been carried out
with both the prolate and oblate deformation of the core (see text for details).
The states with the prolate deformation are found to be dominated by the Nilsson
basis states 1/2[550](= 82$\%$ $\nu$ $h_{11/2}$), 3/2[541](= 88$\%$ $\nu$ $h_{11/2}$),
and 5/2[532](= 93$\%$ $\nu$ $h_{11/2}$) contributing about 50 - 60$\%$, 30$\%$, and
10$\%$, respectively.
The states with the oblate deformation are found to be dominated by the Nilsson
basis states 11/2[505](= 100$\%$ $\nu$ $h_{11/2}$), 9/2[514](= 99$\%$ $\nu$ $h_{11/2}$),
and 7/2[523](= 86$\%$ $\nu$ $h_{11/2}$) contributing about 60 - 85$\%$, 10 - 25$\%$, and
2 - 10$\%$, respectively.
}
\end{figure*}

A comparison of the predictions of the CPCM with the observed
excitation energies is presented in Fig.~\ref{fig:prolate+oblate}.
These calculations were carried out assuming both prolate
and oblate deformation for the $^{102}$Cd core. 
The Nilsson basis states for the underlying levels obtained
in these calculations have been described in the caption of
Fig.~\ref{fig:prolate+oblate}.
 
\begin{figure*}
\includegraphics{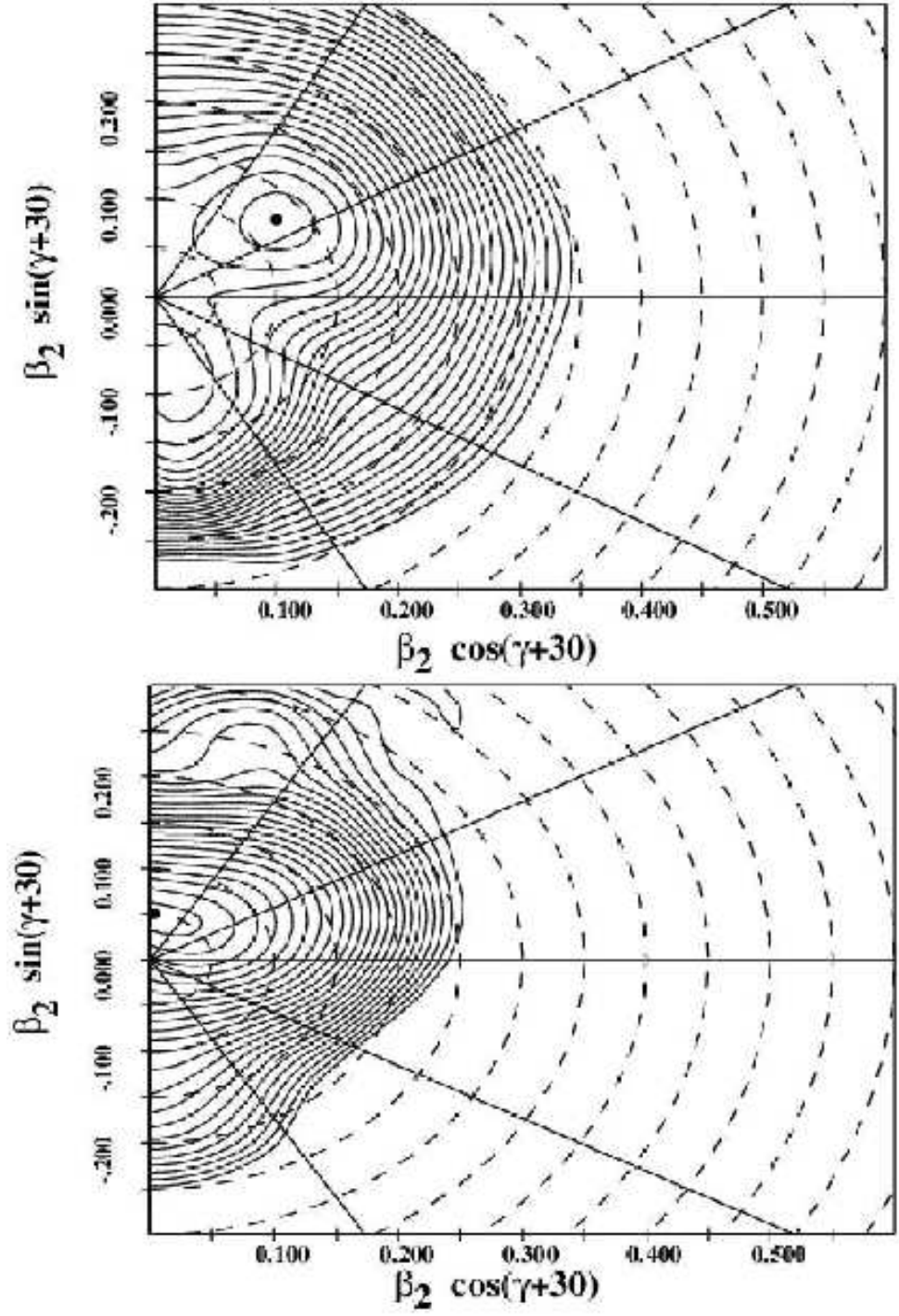}
\caption{
\label{fig:trs}
TRS calculations for states of parity and signature, ($\pi$,$\alpha$) = (-,-1/2),
showing the development of energy minima.
Top panel: $\hbar$$\omega$  = 0.35 MeV; minimum at $\beta_{2}$ = 0.13,
$\gamma$ = 8 $^{\circ}$.
Bottom panel: $\hbar$$\omega$  = 0.70 MeV; minimum at $\beta_{2}$ = 0.05,
$\gamma$ = 57 $^{\circ}$.   
}
\end{figure*}

The theory predicts reasonably well
the experimental spectra up to the 39/2$^{-}$ level for
prolate deformation. 
However, the same could not be reproduced with
the consideration of an oblate deformation of the core.
For higher spins ($>$ 39/2$^{-}$),
the calculation with oblate deformation seems to be in 
better agreement.
Thus, there likely 
is a shape change from prolate to oblate above
the $J^{\pi}$ = 39/2$^{-}$ level at $E_{x}$ = 8011 keV 
in the vicinity of the band termination,
which is corroborated by the Total Routhian Surface (TRS) 
calculations. These calculations also
indicate (see Fig.~\ref{fig:trs}) the evolution
of shape from collective prolate ($\gamma$$\sim$ 10 $^{\circ}$)
to non-collective oblate ($\gamma$$\sim$ 60 $^{\circ}$) 
as the rotational frequency increases.  
Hence the CPCM and TRS calculations both point 
towards the transition from prolate shape to oblate shape
as the excitation energy increases.  

\begin{figure*}
\includegraphics[ height = 18 cm, width = 18 cm,angle = 0]{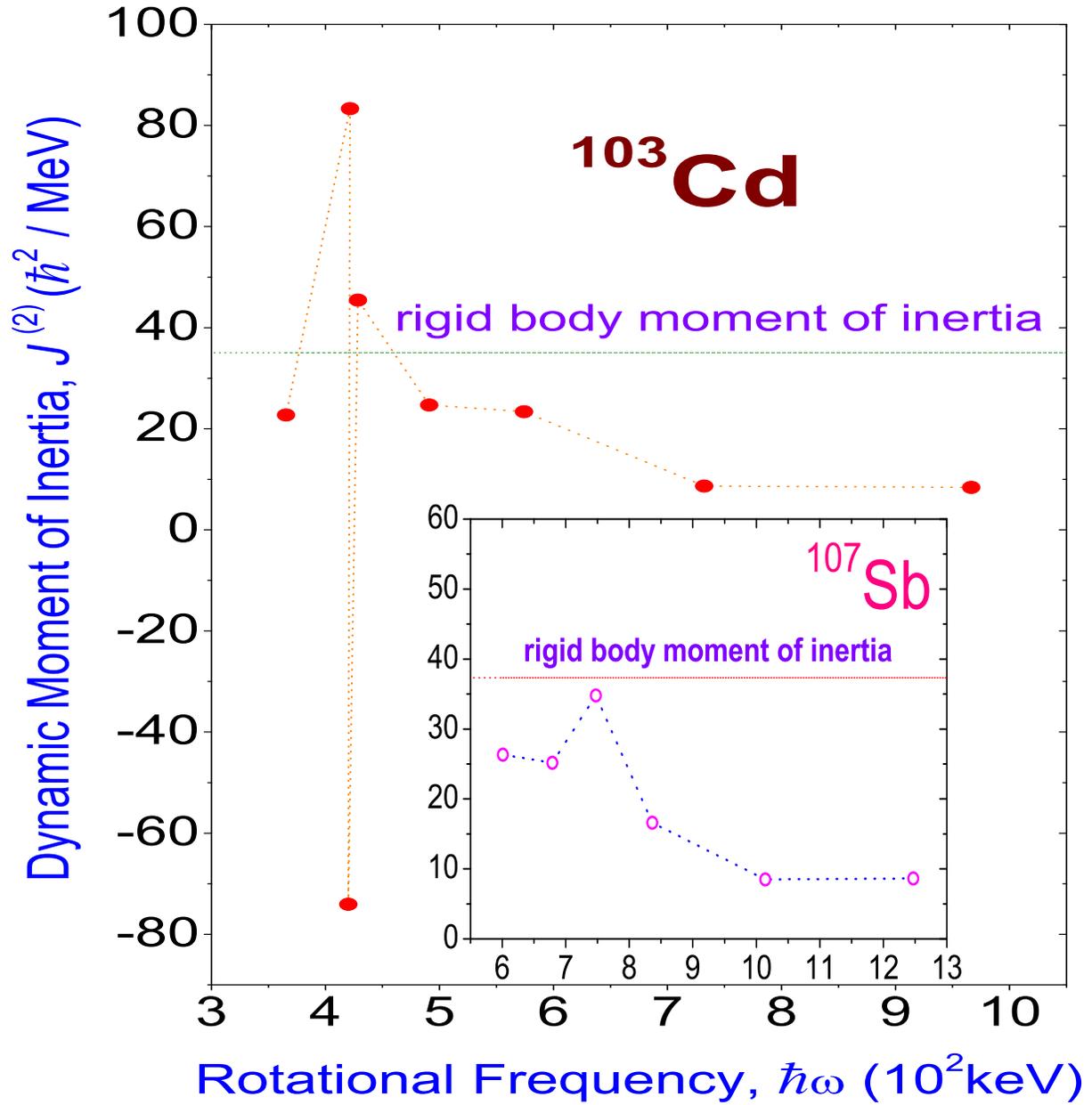}
\caption{
\label{fig:dynamic}
``(Color online)'' The dynamic moment of inertia for $Seq.$ IV in $^{103}$Cd. The dashed line corresponds
to the moment of inertia for a rigid body having mass $A$ = 103 and deformation $\beta$$_{2}$
= 0.23. The inset shows the same for $^{107}$Sb \cite{lafosse}. 
}
\end{figure*}

	To further investigate the terminating nature 	
of the $\Delta$I=2 negative-parity band, we have plotted in 
Fig.~\ref{fig:dynamic} the dynamic moment of inertia
$J$$^{(2)}$ \cite{afan} as a function of rotational frequency
for this band.
In the figure, variation of $J^{(2)}$
with rotational frequency for the $Seq$ IV in $^{103}$Cd 
is compared with that for a rigid rotor with 
$\beta_{2}$ = 0.23 \cite{afan}. The negative value of
$J$$^{(2)}$ at $\hbar\omega$ = 0.42 MeV indicates
backbending, as also discussed in Ref. \cite{palacz}.
Beyond this $\hbar\omega$ value, $J$$^{(2)}$
decreases smoothly with increase in $\hbar\omega$. 
At the top
of the band, $J$$^{(2)}$ is about one fourth of the rigid
rotor value. This is a general feature for the decoupled
smooth terminating bands observed in A = 110 region 
 \cite{lafosse,schnare,wadsworth1,wadsworth2}.
However, a comparison of the observed variation of 
$J$$^{(2)}$ in $^{103}$Cd with the case of other terminating 
bands in the A = 100 \cite{afan, gizon}
and A = 155 - 160 \cite{simpson,afan} regions show a distinct
difference. In contrast to the observed irregular behavior of 
$J$$^{(2)}$ with rotational frequencies for those nuclei, 
the $J$$^{(2)}$ in $^{103}$Cd
appears to be rather smooth. This smooth variation and the decreasing trend
of $J$$^{(2)}$ in $^{103}$Cd is further indicative of a gradual loss of
collectivity and possible change over to oblate shape. It might be surmised, 
that the observed band termination for the
negative-parity sequence corresponds to a ``smooth
termination''.

\section{Conclusions}
The neutron-deficient nucleus $^{103}$Cd has been studied to the high-spin 
regime by means of in-beam $\gamma$-ray spectroscopy using 
Gammasphere. 
More than fifty new transitions have been found and placed in the level scheme.
This has led to a substantial extension of both the positive- and negative-parity
sequences. The decoupled negative parity sequence connected through the cascade
of $E2$ transitions has been observed up to the terminating 47/2$^{-}$ state.
The core particle coupling model 
reproduces both the positive- and negative-parity sequences reasonably well. 
Both phenomenology, CPCM and TRS calculations provide evidence
for a likely transition from prolate to oblate structure
at $J^{\pi} > 39/2^{-}$ 
and for the termination of the negative-parity band;
this termination  
resembles the ``smooth terminating bands'' previously observed in
the $A = 110$ region.  
Further experimental and theoretical work is required to confirm
the smooth terminating nature of the negative parity sequence ($Seq$ IV).
Experimental lifetime information of the levels in $Seq$ IV 
for reliable determination of transition quadrupole moments 
($Q_{t}$ values) would be particularly important to conclude 
firmly about the nature of this sequence.

\section{Acknowledgments}
We wish to thank Drs. M.P. Carpenter, R.V.F. Janssens, B. Kharraja,
F.G. Kondev, and T. Lauritsen for their help with the Gammasphere
experiment and for many fruitful discussions.
The help received from Dr. G. Mukherjee with
the TRS calculation is also gratefully acknowledged.
This work was supported in part
by the National Science Foundation (Grants No. INT-01115336 and PHY04-57120) 
and the University Grants Commission, Government of India.

\newpage
\begin{longtable}{|c|c|c|c|c|c|c|}
\caption{
Gamma transition energies (${E}_{\gamma}$)in keV, initial and final spins, 
relative $\gamma$ ray intensities (${I}_{\gamma}$), DCO-ratios (${R}_{DCO}$),
and multipolarities for the transitions belonging to $^{103}$Cd. 
}
\endfirsthead
\caption[]{continued...} \\  
$Seq$ & $E_{\gamma}$ & $J^{\pi}_{i} \rightarrow J^{\pi}_{f}$ & $I_{\gamma}$ & $R_{DCO}$$^{d}$ & Nature of Gate & Multipolarity \\
    &   (keV)      &                                       &     (Rel.)   &           &                &                \\
\hline
\endhead
\multicolumn{7}{|r|} {continued...~}\\
\endfoot
\hline
\endlastfoot
\hline
$Seq$ & $E_{\gamma}$ & $J^{\pi}_{i} \rightarrow J^{\pi}_{f}$  & $I_{\gamma}$ & $R_{DCO}$$^{d}$ & Nature of Gate & Multipolarity \\
    &   (keV)      &                                        &     (Rel.)   &           &                &                \\
\hline

I  &  118.7(3)  &$21/2^{+}_{1} \ \ \rightarrow 19/2^{+}_{1}$ &   38.3(37)  & 0.53(10)&    $Q$         &  $M1$               \\

    &  170.4(5)  &$25/2^{+}_{1} \ \ \rightarrow 23/2^{+}_{2}$ &   10.2(10)  & 0.52(3)&    $Q$         &  $M1$               \\

    &  187.7(2)  & $7/2^{+}_{1} \ \ \rightarrow 5/2^{+}_{1}$  &   100        & 0.99(11)&    $D$         &  $M1$               \\

    &  208.30(17)  &$(31/2^{+}_{5}) \rightarrow (31/2^{+}_{2})$ &{\it w}$^{a}$ &           &                &                      \\

    &  259.1(2)  &$27/2^{+}_{1} \ \ \rightarrow 25/2^{+}_{1}$ &   32.0(31)  & 0.49(8)&    $Q$         &  $M1$               \\

    &  287.8(3)  &$35/2^{-}_{1} \ \ \rightarrow (35/2^{+}_{1})$ &   2.9(3)   & 0.53(13)& $Q$$^{b}$      &  $E1$               \\

    &  330.6(5)  &$(31/2^{+}_{5}) \rightarrow 31/2^{-}_{1}$ &   5.4(5)   & 0.87(15)& $Q$            &  $E1$               \\

    &  358.7(5)  &                                        &{\it w}$^{a}$ &           &      &                 \\

    &  417.3$^{e}$  & \ \ \ \ \ \ \ \ \ \ \            $\rightarrow (31/2^{+}_{5})$ &{\it w}$^{a}$ &           &                &                 \\

    &  443.7(4)  &$(35/2^{+}_{1}) \rightarrow (31/2^{+}_{5})$ &   9.9(10)   & 0.91(14)& $Q$            &  $E2$               \\

    &  466.60(13)  &$(33/2^{+}_{2}) \rightarrow (31/2^{+}_{3})$ &   7.7(7)   & 0.88(13)& $D$$^{b}$      &  $M1$               \\

    &  537.4(3)  &$(31/2^{+}_{3}) \rightarrow (29/2^{+}_{4})$ &   0.7(1)   & 0.88(16)& $D$$^{b}$      &  $M1$               \\

    &  562.4$^{e}$  & \ \ \ \ \ \ \ \ \ \ \            $\rightarrow (35/2^{+}_{1})$ &{\it w}$^{a}$ &           &                &                 \\

    &  587.1$^{e}$  & \ \ \ \ \ \ \ \ \ \ \             $\rightarrow (31/2^{+}_{5})$ &{\it w}$^{a}$ &           &                &                 \\

    &  623.2(2)  &$19/2^{+}_{1} \ \ \rightarrow 15/2^{+}_{1}$ &   61.2(58)  & 0.99(6)&    $Q$         &  $E2$               \\

    &  642.30(11)  &$(29/2^{+}_{4}) \rightarrow (27/2^{+}_{3})$ &   4.4(5)   & 0.97(13)& $D$$^{b}$      &  $M1$               \\

    &  673.10(14)  &$(31/2^{+}_{5}) \rightarrow (29/2^{+}_{4})$ &   3.7(4)   & 0.89(15)& $D$$^{b}$      &  $M1$               \\

    &  689.2(6)  &$25/2^{+}_{1} \ \ \rightarrow 21/2^{+}_{2}$ &    0.3(1)  &           &      &                 \\

    &  694.6(7)  &$(35/2^{+}_{3}) \rightarrow (33/2^{+}_{2})$ &   8.2(8)   & 0.83(11)& $D$$^{b}$      &  $M1$               \\

    &  704.10(17)  &$(33/2^{+}_{3}) \rightarrow (31/2^{+}_{4})$ &   2.3(3)   & 0.99(14)& $D$$^{b}$      &  $M1$               \\
 
    &  719.7(2)  &$11/2^{+}_{1} \ \ \rightarrow 7/2^{+}_{1}$  &   91.2(87)  & 1.69(5)&    $D$         &  $E2$               \\

    &  796.3(7)  & \ \ \ \ \ \ \ \ \ \ \            $\rightarrow (33/2^{+}_{3})$ &{\it w}$^{a}$ &           &      &                 \\

    &  811.2(5)  &$(29/2^{+}_{2}) \rightarrow 27/2^{+}_{1}$ &   10.8(11)  & 0.76(10)&    $D$         &  $M1$               \\

    &  815.6$^{e}$  &                                        &{\it w}$^{a}$ &           &                &                 \\

    &  841.30(13)  & \ \ \ \ \ \ \ \ \ \ \            $\rightarrow (27/2^{+}_{3})$ &{\it w}$^{a}$ &           &      &                 \\

    &  868.30(25)  &$(31/2^{+}_{3}) \rightarrow (29/2^{+}_{3})$ &   1.8(2)   & 0.78(15)& $D$$^{b}$      &  $M1$               \\
 
    &  920.8(2)  &$15/2^{+}_{1} \ \ \rightarrow 11/2^{+}_{1}$ &   78.7(75)  & 0.94(3)&    $Q$         &  $E2$               \\

    &  942.2(4)  &$(31/2^{+}_{4}) \rightarrow (29/2^{+}_{3})$ &   0.10(2)  & 0.97(18)& $D$$^{b}$      &  $M1$               \\

    &  947.50(17)  &$(31/2^{+}_{5}) \rightarrow (27/2^{+}_{4})$ &   0.7(1)   & 0.60(15)& $Q$$^{b}$      &  $E2$               \\

    &  963.50(11)  &$(27/2^{+}_{3}) \rightarrow 25/2^{+}_{1}$ &   5.0(5)   & 1.04(13)& $D$$^{b}$      &  $M1$               \\

    & 1001.10(18)  &$(31/2^{+}_{2}) \rightarrow (29/2^{+}_{2})$ &   2.7(3)   & 0.88(13)& $D$$^{b}$      & $M1$                \\

    & 1004.00(17)  &$(31/2^{+}_{5}) \rightarrow (29/2^{+}_{3})$ &   1.8(2)   & 1.00(21)&  $D$           & $M1$                \\

    & 1015.90(11)  &$(29/2^{+}_{3}) \rightarrow 27/2^{+}_{1}$ &   5.0(5)   & 0.43(14)&  $Q$           & $M1$                \\

    & 1025.4(4)  &$23/2^{+}_{2} \ \ \rightarrow 21/2^{+}_{1}$ &   14.2(13)  & 1.16(8)&    $D$         & $M1$                \\

    & 1072.90(11)  &$(31/2^{+}_{3}) \rightarrow (29/2^{+}_{2})$ &   4.1(4)   & 0.93(16)& $D$$^{b}$      & $M1$                \\

    & 1108.40(25)  &$(31/2^{+}_{2}) \rightarrow (27/2^{+}_{3})$ &   1.1(1)   &           &      &                 \\

    & 1144.40(16)  &$23/2^{+}_{2} \ \ \rightarrow 19/2^{+}_{1}$ &    2.6(3)  &           &      &                 \\

    & 1146.90(22)  &$(31/2^{+}_{4}) \rightarrow (29/2^{+}_{2})$ &   1.9(2)   & 0.85(16)& $D$$^{b}$      & $M1$                \\
 
    & 1194.5(3)  &$25/2^{+}_{1} \ \ \rightarrow 21/2^{+}_{1}$ &   27.9(26)  & 0.99(6)&    $Q$         & $E2$                 \\

    & 1210.1$^{e}$  & \ \ \ \ \ \ \ \ \ \ \            $\rightarrow (31/2^{+}_{4})$ &{\it w}$^{a}$ &           &       &                 \\

    & 1332.40(10)  &$(27/2^{+}_{4}) \rightarrow 25/2^{+}_{1}$ &   1.3(1)   & 0.86(17)& $D$$^{b}$      & $M1$                \\

    & 1346.70(17)  &$(29/2^{+}_{4}) \rightarrow 27/2^{+}_{1}$ &   1.0(1)   & 1.04(17)& $D$$^{b}$      & $M1$                \\
 
    & 1436.5(3)  & \ \ \ \ \ \ \ \ \ \ \            $\rightarrow (29/2^{+}_{2})$ &   0.10(2)  &           &      &                 \\

    & 1465.3$^{e}$  &$(31/2^{+}_{5}) \rightarrow$              &{\it w}$^{a}$ &           &                &                      \\
 
    & 1539.5(4)  &$(33/2^{+}_{2}) \rightarrow (29/2^{+}_{2})$ &   0.6(1)   & 0.61(18)& $Q$$^{b}$      & $E2$                \\

    & 1608.60(10)  &$(31/2^{+}_{1}) \rightarrow 27/2^{+}_{1}$ &   1.0(1)   & 0.50(18)& $Q$$^{b}$      & $E2$                \\

    & 1712.9$^{e}$  & \ \ \ \ \ \ \ \ \ \ \             $\rightarrow (29/2^{+}_{2})$&{\it w}$^{a}$ &           &                &                 \\

II  & 187.7(6)   &$(19/2^{+}_{3}) \rightarrow 19/2^{+}_{2}$ &    2.3(2)  &           &                &                 \\

    &  449.6(3)  & $(25/2^{+}_{2}) \rightarrow (23/2^{+}_{3})$ &   1.5(2)   &           &                &                 \\

    & 630.70(14)  &                                        &   0.5(1)   &           &                &                 \\

    & 641.1(4)   & \ \ \ \ \ \ \ \ \ \ \            $\rightarrow 19/2^{+}_{2}$ &   5.6(5)   &           &                &                 \\
 
    & 782.4(5)   &$19/2^{+}_{2} \ \ \rightarrow 15/2^{+}_{1}$ &   13.0(12)  & 1.18(9)&      $Q$       &  $E2$           \\

    & 860.1(9)   &$(21/2^{+}_{3}) \rightarrow (19/2^{+}_{3})$ &   8.2(8)   & 0.56(13)&     $Q$        &  $M1$               \\

    &  887.4(6)  &$(25/2^{+}_{2}) \rightarrow (21/2^{+}_{3})$ &   9.1(9)   & 0.95(8)&     $Q$        &  $E2$               \\
 
    & 969.10(11)   &$(19/2^{+}_{3}) \rightarrow 15/2^{+}_{1}$ &    6.4(6)  & 0.98(7)&      $Q$       &  $E2$             \\

    & 1001.7(8)  &$(29/2^{+}_{6}) \rightarrow (25/2^{+}_{2})$ &  10.6(10)   & 0.88(9)&     $Q$        &  $E2$             \\
 
    & 1046.8(9)  &$(21/2^{+}_{3}) \rightarrow 19/2^{+}_{2}$ &   4.2(4)   & 0.53(12)&     $Q$        &  $M1$             \\
 
    & 1099.90(15)  &$(27/2^{+}_{2}) \rightarrow$              &   1.4(1)   &           &                &                 \\

III &   168.4(2)  &$11/2^{+}_{1} \ \ \rightarrow 9/2^{+}_{1}$ &  1.8(2)     & 0.56(19)& $Q$           &   $M1$              \\
 
    &  209.1(4)  &$(23/2^{+}_{1}) \rightarrow 21/2^{+}_{1}$ &   8.0(8)   & 0.52(14)& $Q$            &  $M1$               \\

    & 249.7(7)   &$(31/2^{+}_{2}) \rightarrow (29/2^{+}_{7})$ &   6.0(6)   & 0.81(13)& $D$            & $M1$                \\

    & 256.60(16)   &$(27/2^{+}_{2}) \rightarrow (23/2^{+}_{3})$ &   1.7(1)   & 0.77(14)& $Q$            & $E2$                \\

    & 270.9(8)   &$(35/2^{+}_{1}) \rightarrow (33/2^{+}_{1})$ &   4.6(5)   & 0.91(16)& $D$            & $M1$                \\

    &   318.20(11)  &$15/2^{+}_{1} \ \ \rightarrow 13/2^{+}_{1}$&  9.6(10)     & 0.47(29)& $Q$           &   $M1$              \\

    & 318.6(6)   &$(31/2^{+}_{2}) \rightarrow (29/2^{+}_{5})$ &   4.4(5)   & 0.72(18)& $D$            & $M1$                \\

    &   354.70(15)  &$17/2^{+}_{1} \ \ \rightarrow 15/2^{+}_{1}$&  5.1(5)     & 0.54(23)& $Q$           &   $M1$              \\

    & 380.4(5)   &$(33/2^{+}_{1}) \rightarrow (31/2^{+}_{2})$ &  14.0(13)   & 1.10(26)& $D$            & $M1$                \\

    & 385.30(14)   &$(29/2^{+}_{7}) \rightarrow (27/2^{+}_{5})$ &   3.2(4)   & 0.80(13)& $D$$^{c}$       & $M1$                \\

    & 415.4(5)   &$(35/2^{+}_{2}) \rightarrow (33/2^{+}_{1})$ &   7.7(7)   & 1.05(28)& $D$            & $M1$                \\

    & 425.7(6)   &$(29/2^{+}_{1}) \rightarrow (27/2^{+}_{2})$ &  10.8(10)   & 0.49(11)& $Q$            & $M1$                \\

    & 451.9(9)   &$(31/2)         \rightarrow (29/2^{+}_{1})$ &   4.4(5)   & 0.91(18)& $D$            & $D$                \\

    &   551.9(6)  &$9/2^{+}_{1} \ \ \rightarrow 7/2^{+}_{1}$  &  8.9(9)     & 0.99(5)& $D$           &   $M1$              \\

    &   604.6(6)  &$13/2^{+}_{1} \ \ \rightarrow 11/2^{+}_{1}$& 13.7(13)     & 0.50(7)& $Q$           &   $M1$              \\

    &   672.3(8)  &$17/2^{+}_{1} \ \ \rightarrow 13/2^{+}_{1}$& 10.3(10)     &1.01(6)& $Q$           &   $E2$              \\

    & 739.8(8)   &$(29/2^{+}_{5}) \rightarrow (29/2^{+}_{1})$ &   6.1(6)   & 1.92(27)& $D$            & $M1$                \\

    &   773.1(4)  &$13/2^{+}_{1} \ \ \rightarrow 9/2^{+}_{1}$ & 32.6(31)     & 0.84(4)& $Q$           &   $E2$              \\

    & 808.80(18)   &$(29/2^{+}_{7}) \rightarrow (29/2^{+}_{1})$ &   2.0(2)   & 1.89(22)& $D$            & $M1$                \\

    & 849.30(22)   &$(27/2^{+}_{5}) \rightarrow (27/2^{+}_{2})$ &   3.2(4)   & 1.82(29)& $D$            & $M1$                \\

    &   892.2(8)  &$21/2^{+}_{2} \ \ \rightarrow 17/2^{+}_{1}$& 10.8(10)     &0.98(4)&  $Q$           &   $E2$              \\

    &   947.8(8)  &$19/2^{-}_{1} \ \ \rightarrow 17/2^{+}_{1}$&  4.2(4)     &0.49(13)&  $Q$           &   $E1$              \\

    & 963.4(8)   &$(23/2^{+}_{3}) \rightarrow 19/2^{-}_{1}$ &   5.3(5)   & 1.01(15)& $Q$            & ($M2$)                \\

    &1182.60(24)   &$(29/2^{+}_{1}) \rightarrow 23/2^{+}_{2}$ &   1.1(1)   &           &      &                 \\
 
    & 1572.90(12)  &$(27/2^{+}_{2}) \rightarrow (23/2^{+}_{1})$ &   2.1(2)   & 0.43(14)& $Q$$^{b}$      & $E2$                \\

    &1606.3$^{e}$   &$(27/2^{+}_{5}) \rightarrow 23/2^{+}_{2}$ &{\it w}$^{a}$ &           &                &                 \\

IV    &   566.10(23) & \ \ \ \ \ \ \ \ \ \ \            $\rightarrow 35/2^{-}_{1}$ & 2.3(3)     &           &      &                 \\

      &   580.3(4) & \ \ \ \ \ \ \ \ \ \ \            $\rightarrow 35/2^{-}_{1}$ & 0.60(5)    &           &      &                 \\

      &   598.2(3)  & \ \ \ \ \ \ \ \ \ \ \            $\rightarrow 39/2^{-}_{1}$ & 0.80(7) &           &      &                 \\

      &   609.9(5)     & \ \ \ \ \ \ \ \ \ \ \            $\rightarrow 39/2^{-}_{1}$ & 0.70(8) &           &      &                 \\

      & 643.1(3) &$15/2^{-}_{1} \ \ \rightarrow 11/2^{-}_{1}$  & 22.5(21)   & 0.94(4)&  $Q$             & $E2$                \\

      & 739.9(1) &$9/2^{+}_{1} \ \ \rightarrow 5/2^{+}_{1}$   &  64.7(62)   & 0.88(9)&  $Q$             & $E2$                \\

      &   763.7(4)  & \ \ \ \ \ \ \ \ \ \ \            $\rightarrow 39/2^{-}_{1}$ & 0.60(7) &           &      &                 \\

      & 801.1(4) &$15/2^{-}_{1} \ \ \rightarrow 13/2^{+}_{1}$  & 19.6(19)   & 0.51(21)&  $Q$             & $E1$                \\

      & 813.3(3) &$27/2^{-}_{1} \ \ \rightarrow 23/2^{-}_{1}$  & 37.2(36)   & 0.97(6)&  $Q$             & $E2$                \\
 
      & 818.5(1) &$19/2^{-}_{1} \ \ \rightarrow 15/2^{-}_{1}$ &  41.9(40)   & 1.82(13)&  $D$             & $E2$                \\

      &   849.4(6)  & \ \ \ \ \ \ \ \ \ \ \            $\rightarrow 39/2^{-}_{1}$ & 0.30(4) &           &      &                 \\

      & 867.4(3) &$23/2^{-}_{1} \ \ \rightarrow 19/2^{-}_{1}$  & 36.0(34)   & 0.93(5)&  $Q$             & $E2$                \\

      & 901.4(3) &$31/2^{-}_{1}  \ \ \rightarrow 27/2^{-}_{1}$  & 26.0(25)   & 0.88(14)&  $Q$             & $E2$                \\

      &   917.2(7) &                                        & 1.2(2)     &           &      &                 \\

      &  923.3(2)&$23/2^{-}_{1} \ \ \rightarrow 21/2^{+}_{2}$  &  0.8(1)   &           &      &                 \\
 
      & 930.6(3) &$11/2^{-}_{1} \ \ \rightarrow 9/2^{+}_{1}$  & 25.9(25)    & 0.50(14)&  $Q$             & $E1$                \\ 

      &  1012.4(4) & \ \ \ \ \ \ \ \ \ \ \            $\rightarrow 35/2^{-}_{1}$  & 0.30(4)   &           &      &                 \\

      &1057.4(6) & \ \ \ \ \ \ \ \ \ \ \                $\rightarrow 31/2^{-}_{1}$  &  0.20(3)  &          &      &                 \\
 
      &1063.4(4) &$35/2^{-}_{1} \ \  \rightarrow 31/2^{-}_{1}$  & 12.9(12)   & 0.81(18)&  $Q$             &$E2$                 \\

      &  1131.5(5)  &                                        & 0.30(4) &           &      &                 \\

      &   1151.4(7)  & \ \ \ \ \ \ \ \ \ \ \            $\rightarrow 39/2^{-}_{1}$ & 0.50(4)&          &      &                 \\
 
      &1234.2(7) &$39/2^{-}_{1} \ \ \rightarrow 35/2^{-}_{1}$  &  4.1(4)   & 0.71(17)&  $Q$             &$E2$                 \\
 
      &1429.90(14)&$23/2^{-}_{1} \ \ \rightarrow 21/2^{+}_{1}$  &  1.5(2)   &           &      &                 \\

      &  1675.1(3)  & \ \ \ \ \ \ \ \ \ \ \            $\rightarrow 39/2^{-}_{1}$ & 0.40(4) &           &      &                 \\

      &  1695.4(6)  &$(43/2^{-}_{1})\rightarrow 39/2^{-}_{1}$ & 0.40(4)&          &      & ($E2$)          \\

      &   1713.7(5)  & \ \ \ \ \ \ \ \ \ \ \            $\rightarrow 39/2^{-}_{1}$ & 0.30(4)&          &      &                 \\
 
      &  1778.4(3)  & \ \ \ \ \ \ \ \ \ \ \            $\rightarrow 35/2^{-}_{1}$ & 0.40(5) &           &      &                 \\

      &  2170.6$^{e}$  &$(47/2^{-}_{1})\rightarrow (43/2^{-}_{1})$ &{\it w}$^{a}$&         &      &  ($E2$)         \\

\end{longtable}
\footnotetext[1] {a: Corresponds to weak transition whose intensity could not be computed.}
\footnotetext[2] {b: The transition of interest is used as the gating transition and the cited DCO-value corresponds to the known 119-keV transition. Multipolarity of the gating transition is determined in this process.}
\footnotetext[3] {c: The transition of interest is used as the gating transition and the cited DCO-value corresponds to the known 250-keV transition. Multipolarity of the gating transition is determined in this process.}
\footnotetext[4] {d: A blank is left for those transitions for which the $R_{DCO}$ could not be computed.} 
\footnotetext[5] {e: The peak could not be fitted due to poor statistics. The peak energy was determined from
the centroid position of the peak and hence the corresponding fitting error could not be computed.} 
                                                                                                   
\newpage

\begin{table}
\caption{
Parameters used for the CPCM calculation for the positive parity levels
}
\begin{ruledtabular}
\begin{tabular}{ccccccc}
Even - even core  & Deformation & $\mu$ & $\kappa$ & $\lambda$ & $\Delta$ & Attenuation Factor for \\
                  & $\delta$(= 0.95 $\beta_{2}$) &  &  &  &             & the Coriolis Matrix Elements   \\
 
\hline
 
$^{102}$Cd & 0.14\footnotemark[1] & 0.22\footnotemark[2] & 0.06\footnotemark[2] & 49.0 MeV & 1.2 MeV & 0.81   \\
 
\footnotetext[1]  {Taken from the Ref.\cite{palacz}}
\footnotetext[2]  {Taken from the Ref.\cite{meyer1}}
 
\end{tabular}
\end{ruledtabular}
 
\end{table}

\begin{longtable}{|c|c|c|c|c|c|}
\caption{
Comparison of the experimental and theoretical
branchings for some of the positive-parity levels.
The excitation energies have been expressed in keV.
The branchings have been expressed in $\%$.
The extreme left column denotes the sequence
to which the initial decay state belongs.}
\endfirsthead
\caption[]{continued...} \\
$Seq$ & $J_{i}$ & $J_{f}$   & $E_{{x}_{i}}$     &    $E_{{x}_{f}}$        & Branching   \\
    &               &                 & Expt.    (Theo.)  &  Expt.      (Theo.)     &  Expt.     (Theo.)\\
\hline
\endhead
\multicolumn{6}{|r|} {continued...~}\\
\endfoot
\hline
\endlastfoot
\hline
$Seq$ & $J_{i}$ & $J_{f}$   & $E_{{x}_{i}}$       &    $E_{{x}_{f}}$          & Branching   \\
    &               &                 & Expt.    (Theo.)    &  Expt.      (Theo.)       &  Expt.     [Theo.]\\
\hline

I  & 11/2$_{1}$  & 7/2$_{1}$         &908       (1088)     &  188          (127)       &  98.1$\pm$13.1        [99.9]       \\
    & 11/2$_{1}$  & 9/2$_{1}$         &908       (1088)     &  740          (924)       &   1.9$\pm$0.3        [ 0.1]       \\
\hline

I  & 15/2$_{1}$  & 11/2$_{1}$        &1829      (2089)     &  908         (1088)       &  89.1$\pm$11.4        [99.9]       \\
    & 15/2$_{1}$  & 13/2$_{1}$        &1829      (2089)     &  1513        (1724)       &  10.9$\pm$1.5        [0.1]       \\
\hline

I  & 23/2$_{2}$  & 21/2$_{1}$        &3596      (3779)     &  2571        (3165)       &  84.5$\pm$10.2        [97.4]       \\
    & 23/2$_{2}$  & 19/2$_{1}$        &3596      (3779)     &  2452        (2645)       &  15.5$\pm$2.2        [2.6]       \\
\hline
 
I  & 25/2$_{1}$  & 21/2$_{1}$        &3766      (3849)     &  2571       (3165)        &  72.7$\pm$8.6        [99.2]       \\
    & 25/2$_{1}$  & 23/2$_{2}$        &3766      (3849)     &  3596        (3779)       &  26.6$\pm$3.2        [0.02]       \\
    & 25/2$_{1}$  & 21/2$_{2}$        &3766      (3849)     &  3077        (3324)       &   0.8$\pm$0.3        [0.8]       \\
\hline

I  & 29/2$_{4}$  & 27/2$_{1}$        &5372      (5733)     &  4025        (4505)       &  18.5$\pm$2.5        [41.7]       \\
    & 29/2$_{4}$  & 27/2$_{3}$        &5372      (5733)     &  4730        (5033)       &  81.5$\pm$12.0        [58.3]       \\
\hline

I  & 31/2$_{2}$  & 27/2$_{3}$        &5837      (5684)     &  4730        (5033)       &   7.7$\pm$0.8        [0.2]       \\
    & 31/2$_{2}$  & 29/2$_{2}$        &5837      (5684)     &  4836        (5287)       &  19.0$\pm$2.4        [99.4]       \\
    & 31/2$_{2}$  & 29/2$_{5}$        &5837      (5684)     &  5519        (5785)       &  31.0$\pm$4.0        [0.3]       \\
    & 31/2$_{2}$  & 29/2$_{7}$        &5837      (5684)     &  5588        (6032)       &  42.3$\pm$4.9        [0.1]       \\
\hline

I  & 31/2$_{3}$  & 29/2$_{2}$        &5909      (6094)     &  4836        (5287)       &  62.1$\pm$7.4        [95.2]       \\
    & 31/2$_{3}$  & 29/2$_{4}$        &5909      (6094)     &  5372        (5733)       &  10.6$\pm$1.7        [0.12]       \\
    & 31/2$_{3}$  & 29/2$_{3}$        &5909      (6094)     &  5041        (5411)       &  27.3$\pm$3.6        [4.7]       \\
\hline

I  & 31/2$_{4}$  & 29/2$_{2}$        &5983      (6370)     &  4836        (5287)       &  95.0$\pm$13.8        [74.4]       \\
    & 31/2$_{4}$  & 29/2$_{3}$        &5983      (6370)     &  5041        (5411)       &   5.0$\pm$1.1        [25.6]       \\
\hline

I  & 33/2$_{2}$  & 29/2$_{2}$        &6376      (6264)     &  4836        (5287)       &  7.2$\pm$1.4         [99.9]       \\
    & 33/2$_{2}$  & 31/2$_{3}$        &6376      (6264)     &  5909        (6094)       &  92.8$\pm$11.6        [0.1]       \\     
\hline

I  & 35/2$_{1}$  & 31/2$_{5}$        &6489      (6398)     &  6045        (6446)       &  68.3$\pm$8.7        [26.9]       \\
    & 35/2$_{1}$  & 33/2$_{1}$        &6489      (6398)     &  6218        (5548)       &  31.7$\pm$4.2        [73.1]       \\
\hline

II   & 19/2$_{3}$  & 15/2$_{1}$        &2799      (3088)     &  1829        (2089)       &  73.6$\pm$8.7        [97.1]       \\
    & 19/2$_{3}$  & 19/2$_{2}$        &2799      (3088)     &  2611        (2986)       &  26.4$\pm$3.0        [2.9]       \\
\hline

II   & 21/2$_{3}$  & 19/2$_{2}$        &3658      (3393)     &  2611        (2986)       &  33.9$\pm$4.0        [99.6]       \\
    & 21/2$_{3}$  & 19/2$_{3}$        &3658      (3393)     &  2799        (3088)       &  66.1$\pm$8.0        [0.4]       \\
\hline

II   & 25/2$_{2}$  & 21/2$_{3}$        &4545      (4335)     &  3658        (3393)       &  85.8$\pm$11.3        [68.7]       \\
    & 25/2$_{2}$  & 23/2$_{3}$        &4545      (4335)     &  4096        (4092)       &  14.2$\pm$2.3        [31.3]       \\
\hline

III & 9/2$_{1}$   & 5/2$_{1}$         &740       (924)      &  0           (0)          &  87.9$\pm$11.3        [92.4]       \\
    & 9/2$_{1}$   & 7/2$_{1}$         &740       (924)      &  188         (127)        &  12.1$\pm$1.6        [7.6]       \\
\hline

III & 13/2$_{1}$  & 9/2$_{1}$         &1513      (1724)     &  740         (924)        &  70.4$\pm$8.4        [66.4]       \\
    & 13/2$_{1}$  & 11/2$_{1}$        &1513      (1724)     &  908         (1088)       &  29.6$\pm$3.5        [33.6]       \\
\hline
 
III & 17/2$_{1}$  & 13/2$_{1}$        &2185      (2201)     &  1513        (1724)       &  66.9$\pm$8.1        [96.9]       \\
    & 17/2$_{1}$  & 15/2$_{1}$        &2185      (2201)     &  1829        (2089)       &  33.1$\pm$4.0        [3.1]       \\
\hline

III & 29/2$_{7}$  & 29/2$_{1}$        &5588      (6032)     &  4779        (4666)       &  38.5$\pm$5.1        [90.0]       \\
    & 29/2$_{7}$  & 27/2$_{5}$        &5588      (6032)     &  5202        (5368)       &  61.5$\pm$9.4        [10.0]       \\
\hline

\end{longtable}

\end{document}